\documentclass[twocolumn,showpacs,preprintnumbers,amsmath,amssymb]{revtex4}

\usepackage{graphicx}
\usepackage{dcolumn}
\usepackage{bm}
\usepackage{epsfig}

\begin{document}

\title{Generalized exclusion statistics and degenerate signature of strongly interacting anyons}
\author{M.T. Batchelor and X.-W. Guan}
\affiliation{Department of Theoretical Physics, Research School of Physical Sciences and Engineering, and\\
Mathematical Sciences Institute,
Australian National University, Canberra ACT 0200,  Australia}

\date{\today}

\begin{abstract}
\noindent
We show that below the degenerate temperature the distribution profiles of strongly interacting anyons 
in one dimension coincide with the most probable distributions of ideal particles obeying 
generalized exclusion statistics (GES).  
In the strongly interacting regime 
the thermodynamics and the local two-particle correlation function derived from the GES are seen to agree 
for low temperatures 
with the results derived for the anyon model using the thermodynamic Bethe Ansatz. 
The anyonic and dynamical interactions implement a continuous 
range of GES, providing a signature of strongly interacting anyons, 
including the strongly interacting one-dimensional  Bose gas.
\end{abstract}



\maketitle
\section{Introduction}
\label{Intro}

A fundamental principle of quantum statistical mechanics is the existence of two types of particles -- 
bosons and fermions -- obeying Bose-Einstein and Fermi-Dirac statistics.
The Pauli exclusion principle dictates that no two fermions can occupy the same quantum state.
There is no such restriction on  bosons. 
These rules lead to a number of striking and subtle quantum many-body effects.
In three dimensions particles are either bosons or fermions.
However, in two-dimensions, anyons \cite{F-S} may also exist, obeying fractional statistics.
Fractional statistics have recently been observed in experiments on the elementary excitations of 
a two-dimensional electron gas in the fractional quantum Hall (FQH) regime
\cite{cond1}, where the quasi-particles are charged anyons.
The concept of anyons has become important in studying the FQH effect
\cite{cond1,QFHE,Wilczek} and in topological quantum computation \cite{Wilczek,comp}.

A more general description of quantum statistics is provided by
Haldane exclusion statistics \cite{Haldane}. 
Haldane formulated a description of fractional statistics based on a
generalized Pauli exclusion principle, which is now called generalized
exclusion statistics (GES).
The definition of GES is independent of space dimension. A key point
of GES is to count the dimensionality of the single-particle
Hilbert space as extra particles are added.
It is of interest to find applications of Haldane GES in
one-dimensional (1D) and two-dimensional quantum systems.
A particular property of 1D many-body interacting systems
is that pairwise dynamical interaction between identical particles is
inextricably related to their statistical interaction.
Anyons in one dimension acquire a step-function-like phase when two
identical particles exchange their positions in the scattering process. 
It follows that 1D topological effects and dynamical interaction are not separable.
This transmutation between dynamical interaction and statistical
interaction has been observed in explicit calculations on the 1D
Calogero-Sutherland model \cite{Ha} and the 1D interacting Bose gas
\cite{Wu,Isakov}, which are equivalent to an ideal gas obeying GES.
The 1D interacting model of anyons proposed by Kundu \cite{Kundu}
provides further evidence of the transmutation between dynamical and
statistical interaction \cite{BGO}.
This integrable anyon model reduces to the 1D interacting Bose gas
\cite{LL} and the free Fermi gas as special cases. 
Very recently an anyon-fermion mapping has been proposed to obtain 
solutions for several models of ultracold gases with $1$D anyonic exchange symmetry \cite{Girardeau}.

The experimental realization of the 1D 
Tonks-Girardeau (TG) gas \cite{T-G1, T-G2,T-G3}, has provided significant insights into the
fermionization of interacting Bose systems.
The pure TG gas is obtained in the infinite interaction strength limit of the 1D Bose gas \cite{LL}.
The experimental measurement of local pair correlations in the strongly interacting 1D Bose
gas reveals an important feature, namely that the wave functions overlap even in the 
TG regime \cite{Weiss}.
This implies that the strongly interacting 1D Bose gas does not exhibit pure Fermi behaviour.
Equivalently, the statistics are not strictly Fermi-Dirac.
A deviation from Fermi-Dirac statistics also appears in the strongly interacting limit of the
anyon gas \cite{BGO}, also called the anyonic TG gas \cite{Girardeau}.

In this paper we give a quantitative description of the distribution profiles
of the strongly interacting 1D anyon gas, 
including the strongly interacting Bose gas and the TG limit.
We show that these profiles are equivalent to the most probable distribution profiles of ideal particles 
obeying nonmutual GES at temperatures below a degenerate temperature $T_d$.
This equivalence between strongly interacting anyons and ideal particles
with nonmutual GES is further supported  by the thermodynamic properties
and the local two-body correlation functions, which we derive independently 
from the thermodynamic Bethe Ansatz (TBA) equations for the anyon model 
and via GES \cite{Isakov2,Wilczek-Sen,Iguchi}.
We conclude that 1D interacting anyons and bosons in the
strong coupling regime are properly described by GES. The
quasiparticle excitations for these systems obey GES. The anyonic
interaction and the dynamical interaction implement a continuous range
of GES.  At low temperatures, i.e., $k_BT< 0.1 T_d$, the dynamical
interaction and anyonic statistical interaction determine the GES and
the thermal fluctuations are suppressed. At zero temperature they
recover the properties derived from the Bethe ansatz equations (BAE)
\cite{BGO}.
In addition, the distribution profiles indicate that the dynamical
interaction and anyonic statistical interaction may implement a
duality between the GES parameter $\alpha$ and its inverse $1/\alpha$.

This paper is set out as follows. In section \ref{model}, we present
the Bethe Ansatz solution of the 1D interacting anyon model. The
ground state properties are also analysed. We demonstrate the quivalence
between 1D interacting anyons and ideal particles obeying GES in
section \ref{GES}. Distribution profiles of 1D interacting
anyons and ideal particles obeying GES are discussed in section
\ref{distribution}. In section \ref{TD}, we compare the thermodynamics and the
local pair correlation functions obtained via the TBA and GES approaches. Section
\ref{conclusion} is devoted to concluding remarks and a brief discussion.

\section{The model}
\label{model}

We consider the 1D interacting anyon model with hamiltonian \cite{Kundu}
\begin{equation}
H_N=-\frac{\hbar ^2}{2m}\sum_{i = 1}^{N}\frac{\partial
^2}{\partial x_i^2}+\,g_{\rm 1D}\sum_{1\leq i<j\leq N} \delta
(x_i-x_j)\label{Ham1}
\end{equation}
under periodic boundary conditions. 
Here $m$ denotes the atomic mass.
In analogy with the 1D confined Bose gas \cite{Olshanii}, the coupling constant is determined by 
$g_{\rm 1D}={\hbar^2c}/{m}=-{2\hbar ^2}/{(ma_{\rm 1D})}$ where the coupling strength $c$ is tuned 
through an effective $1$D scattering length $a_{\rm 1D}$ via confinement.
Hereafter we set $\hbar =2m=1$ for convenience. 
We shall restore physical units in the thermodynamics later.
The ground state properties and
Haldane exclusion statistics for the model (\ref{Ham1}) have been
studied recently in Ref.~\cite{BGO}.
The hamiltonian (\ref{Ham1}) exhibits both anyonic statistical and dynamical interactions 
which result in a richer range of quantum effects than those of the 1D interacting Bose gas \cite{LL}.

The Bethe Ansatz wave function is written as \cite{Kundu,BGO}
\begin{eqnarray}
\chi(x_1\ldots
x_N)&=&{\mathrm e}^{-\frac{\mathrm{i}\kappa}{2}\sum_{x_i<x_j}^Nw(x_i,x_j)}\sum_PA(k_{P1}\ldots
k_{PN})\nonumber\\
& & \times \, {\mathrm e}^{\mathrm{i}(k_{P1}x_1+\cdots +k_{PN}x_N)}.
\label{wave}
\end{eqnarray}
It satisfies the anyonic symmetry 
\begin{equation}
\chi(\ldots x_i\ldots x_j \ldots)={\mathrm e}^{-\mathrm{i}\theta } \chi(\ldots x_j\ldots x_i \ldots )
\end{equation}
with the anyonic phase 
\begin{equation}
\theta = \kappa
\left(\sum_{k=i+1}^jw(x_i,x_k)-\sum_{k=i+1}^{j-1}w(x_j,x_k) \right)
\end{equation}
for $i<j$.  
In equation (\ref{wave}), the summation is over all $N!$
permutations $P$. The coefficents $A(k_{P1}\ldots k_{PN})$ are
determined by the two-body scattering relation  \cite{Kundu,BGO}
\begin{equation}
A(\ldots k_j,k_i
\ldots)=\frac{k_j-k_i+\mathrm{i}c'}{k_j-k_i-\mathrm{i}c'}A(\ldots
k_i,k_j \ldots).
\end{equation}
In the above equations the multi-step
function $w(x_1,x_2)=-w(x_2,x_1)=1$ for $x_1>x_2$, with $w(x,x)=0$.
$\kappa$ is the anyonic interaction parameter.  $\left\{x_i\right\}$
are the coordinates of particles in length $L$.  
The anyonic parameter $\kappa$ is the 
key difference from the standard $1$D Bose gas \cite{LL}, for which $\kappa=0$.  
The interacting
anyons in the strong coupling regime may be viewed as the anyonic TG
gas with  zero range interaction. This has been  verified  using 
anyon-fermion mapping for the hard core anyons \cite{Girardeau}.

The energy eigenvalues are given by $E=\sum_{j=1}^N k_j^2$, where the
individual quasimomenta $k_j$ satisfy \cite{Kundu,BGO}
\begin{equation}
{\mathrm e}^{\mathrm{i}k_jL}=-{\mathrm e}^{\mathrm{i}\kappa(N-1)} \prod^N_{\ell = 1} 
\frac{k_j-k_\ell+\mathrm{i}\, c'}{k_j-k_\ell-\mathrm{i}\, c'} 
\label{BA}
\end{equation}
for $j = 1,\ldots, N$.  We will refer to these equations as the BAE.
The anyonic parameter $\kappa$ and the dynamical interaction $c$
are inextricably related via the effective coupling constant 
\begin{equation}
c'=c/\cos(\kappa/2).
\end{equation}

We use a dimensionless coupling constant $\gamma =c/n$ to characterize different physical 
regimes of the anyon gas, where $n=N/L$ is the linear density. 
In this paper, we consider the TG regime where $\gamma \gg 1$ with $0 \le \kappa <4\pi$. 
The model reduces to the interacting Bose gas \cite{LL} at $\kappa=0$.
For $\kappa=\pi$ and $3\pi$ it reduces to the free Fermi gas.  
The anyonic TG gas lies in the range $0\le \kappa \le \pi$ and $3\pi \le \kappa \le 4\pi$, where
the effective interaction $c'>0$.  
However, if  the anyon parameter $\kappa$ is tuned smoothly from $\kappa <\pi$ to $\kappa>\pi$, 
i.e.,  $\pi \le\kappa \le 3\pi$, the effective interaction is attractive. 
Here the strong Fermi pressure may prevent the anyons from collapsing.  
We call this the super anyonic TG gas.

In general the global phase factor behaves like a topological effect associated with Aharonov-Bohm and
Aharonov-Casher fluxes in a 1D mesoscopic ring \cite{Zhu}. 
The parameter $\kappa$ leads to a shift in quasimomenta $k_i\to k_i+\kappa(N-1)/L$ which gives rise to
excited states.  
Due to the periodicity of the BAE, we only consider an effect from
$\kappa(N-1) = \nu$ (mod $2\pi$) with $-\pi \leq \nu \leq \pi$
\cite{BGO}. In the thermodynamic limit, the ground state is given by
$E=N(n^2e(\gamma,\kappa)+\nu^2/L^2)$ where
$e(\gamma,\kappa)=\frac{\gamma^3}{\lambda^3}\int_{-1}^1g(x)x^2dx$.
The  root density distribution $g(x)$ and the parameter $\lambda =c/Q$ are
determined by Lieb-Liniger type integral  equations of the form
\begin{eqnarray}
 g(x)&=&\frac{1}{2\pi}+\frac{\lambda \cos({\kappa}/{2})
  }{\pi}\int_{-1}^{1}\frac{g(y)dy}{\lambda^2+\cos^2(\frac{\kappa}{2})(x-y)^2} \nonumber\\
  \lambda&=& \gamma \int_{-1}^1g(x)dx. 
\end{eqnarray}
In the above $Q$ is the cut-off momentum. 
For the strong coupling regime
the lowest energy per particle is given by
\begin{equation}
\frac{E_0}{N} \approx
\frac{\hbar ^2}{2m}\frac{\pi^2}{3}n^2\left(1-{4 \gamma^{-1}
\cos({\kappa}/{2})}\right) + {\nu^2}/{L^2}.
\end{equation}

In the thermodynamic limit, the last term can be ignored
compared to the kinetic energy.  At zero temperature the ground state
energy $E_0$ can be identified with that for hard-rod anyons
with an effective length $L_{\rm eff} = L(1-na_{1D}\cos (\kappa/2))$
for $\gamma=-2/(na_{1D}) \gg 1$. Here  $a_{1D}$ is negative. The effective
length becomes smaller than $L$.  This causes the linear density $n$ to be 
increased, thus the kinetic energy, being proportional to $n^2$, also increases. 
However, if $0< \kappa < \pi$ or $3\pi< \kappa < 4\pi$, the
effective length becomes larger than $L$ due to repulsion. This
leads to a smaller kinetic energy than the free Fermi energy. 
In this way the anyonic paramter $\kappa$ triggers a resonance-like behaviour from
strongly repulsive interaction to strongly attractive interaction. 
In the resonance region, i.e.,  $na_{1D}\cos (\kappa/2)) \approx 0$, 
the TG anyon gas and the super TG gas correspond to the
GES duality between $\alpha $ and its inverse $1/ \alpha$. 
We define the GES parameter $\alpha$ and discuss this behaviour further below.

\section{GES description}
\label{GES}

According to Haldane exclusion statistics, the
GES parameter $\alpha_{ij}$ is defined through the linear relation
${\Delta d_i}/{\Delta N_j}=-\alpha _{ij}$ \cite{Haldane}, i.e., the
number of available single particle states of species $i$, denoted by
$d_i$, depends on the number of other species $\left\{N_j\right\}$ when 
one particle of species $i$ is added.  Thus $d_i$ is given by \cite{Wu,Isakov}
\begin{equation}
d_i(\left\{N_j\right\})=G_i^0-\sum_j\alpha
_{ij}N_j.
\end{equation}
Here $G_i^0 = d(\left\{0 \right\})$ is the number of available single
particle states when no particles present in the system.
The number of configurations at fixed
$\left\{N_j\right\}$ when one particle of the $i$th species is excited is \cite{Wu}
\begin{equation}
W(\left\{N_i\right\})=\prod_i\frac{(d_i+N_i-1)!}{(N_i)!(d_i-1)!}
\end{equation}
which gives the grand partition function
\begin{equation}
Z=\sum_{\left\{ N_i\right\}}W(\left\{N_i\right\}) {\mathrm e}^{\sum_iN_i(\mu_i-\epsilon_i)/K_BT}.
\end{equation}

In deriving the quantum statistics \cite{Wu},  
the most probable distribution is given by $\sum _j(w_j\delta_{ij}+\beta_{ij})n_j=1$, where 
$w_i$ is determined  by
\begin{equation}
(1+w_i)\prod _j\left(\frac{w_j}{1+w_j}\right)^{\alpha  _{ji}}={\mathrm e}^{(\epsilon_i-\mu_i)/K_B T}.
\label{mutual}
\end{equation}
In the above, $n_i = N_i/G_i$ and $\beta _{ij} = \alpha_{ij} G_j/G_i$ as defined in Ref.~\cite{Wu}.
For $\alpha=0$ the GES result (\ref{mutual}) reduces to Bose-Einstein statistics with
$n_i = 1/({\mathrm e}^{(\epsilon_i-\mu)/K_B T} - 1)$.
For $\alpha_{ij}=1$ it reduces to Fermi-Dirac statistics with
$n_i = 1/({\mathrm e}^{(\epsilon_i-\mu)/K_B T} + 1)$.

On setting $w_i={\mathrm e}^{\epsilon(k_i)/K_BT}$ we see that equation (\ref{mutual}) becomes
\begin{equation}
\epsilon(k_i)=\epsilon_i-\mu_i-K_BT\sum_j
\left(\delta_{ji}-\alpha_{ji}\right)\left(1+{\mathrm e}^{-\frac{\epsilon(k_j)}{K_B T}}\right)
\end{equation}
which  is identical to the TBA  equations 
\begin{equation}
\epsilon (k)=
\epsilon -\mu-\frac{K_BT}{2\pi}\int_{-\infty}^{\infty}dk^{'}\theta
^{'}(k-k^{'})\ln(1+{\mathrm e}^{-\frac{\epsilon(k^{'})}{K_B T}}) \label{TBA}
\end{equation}
for the anyon model subject to the identification 
\begin{equation}
 \alpha _{ij}
:=\alpha (k,k') =\delta(k,k')-\frac{1}{2\pi}\theta'(k-k')
\end{equation}
with a
dispersion relation $\epsilon=\frac{\hbar^2}{2m}k^2$. This
definition allows different species to refer to identical particles
with different quantum numbers or with different momenta.
Here the function
\begin{equation}
\theta'(x)=\frac{2c\cos(\kappa/2)}{c^2+\cos^2(\kappa/2)x^2}.
\end{equation}
We note that this type of connection between TBA and GES appears to be very general
for 1D interacting systems \cite{Wadati}, including multicomponent
systems \cite{Sutherland}. 

For the TBA, quasiparticle excitations are characterized  by the density of
occupied states and the density of holes denoted by $\rho$ and
$\rho_{\rm h}$, repectively.  The dressed energy is expressed by
$\rho_{\rm h}/\rho={\mathrm e}^{\frac{\epsilon(k)}{K_B T}}$. 
In the ground state $\rho_{\rm h}=0$.  
At  arbitrary temperatures, we have the BAE
\begin{equation}
\rho(k)+\rho_{\rm h}(k)=\frac{1}{2\pi}+\frac{1}{2\pi}\int
_{-\infty}^{\infty}dk'\theta'(k-k')\rho(k') \label{d-BA}
\end{equation} 
which can give rise in the strongly interacting limit to a relation between the particle distribution $\rho(k)$
and the hole distribution $\rho_{\rm h}(k)$ of the form
\begin{equation}
2\pi\left(\rho(k)+\rho_{\rm h}(k)\right)\approx 1/\alpha
\end{equation}
with \cite{BGO}
\begin{equation}
\alpha = 1- {2 \gamma^{-1} \cos(\kappa/2)}. 
\end{equation}
This relation clearly indicates the nature of the deviation from Fermi-Dirac statistics. 
At low temperatures $\rho_{\rm h} \ll \rho$. 
It follows that, up to the negligible quantity $2\cos(\kappa/2)\rho_{\rm h}/(\gamma \rho)$, we have 
$2\pi\left(\alpha \rho +\rho_{\rm h}\right)\approx 1$.  
This relation indicates that at low temperatures a one particle excitation leaves $\alpha$ holes below the Fermi surface. 
Minimizing the free energy  \cite{Ha}
\begin{eqnarray}
& &\frac{F}{L}=\int_{-\infty}^{\infty} dk\rho \left(\epsilon -\mu\right)\\
& &-K_BT\int_{-\infty}^{\infty} dk\left\{\left(\rho+\rho_{\rm h}
\right)\ln ( \rho+\rho_{\rm h})-\rho \ln \rho-\rho_{\rm h}\ln
\rho_{\rm h}\right\}\nonumber
\end{eqnarray}
with this relation gives the most probable distribution $n(\epsilon)=2\pi \rho$ 
to be
\begin{equation}
n(\epsilon)=\frac{1}{\alpha+ w(\epsilon)} \label{FS}
\end{equation}
where  $w(\epsilon)$ satisfies the equation
\begin{equation}
w^{\alpha}(\epsilon)\left(1+w(\epsilon)\right)^{1-\alpha}={\mathrm e}^{\frac{\epsilon-\mu}{K_BT}}.\label{FS-2}
\end{equation} 
This is precisely equation (\ref{mutual}) with nonmutual statistics,  i.e., with  $\alpha_{ij}=\alpha \delta _{ij}$.

We have thus shown that strongly interacting anyons map to an ideal gas obeying nonmutual GES $\alpha$ at low temperatures.
We derive the statistics from (\ref{FS}) below.
It is clear to see that at zero temperature there is a Fermi-like
 surface with a cut-off energy $\epsilon_{\rm F}=\mu_0$, where $n(k)=
 1/\alpha $ if $\epsilon(k)\leq \mu_0$ and $n(k)=0$ if $\epsilon(k) > \mu_0$. 
 At low temperatures, a relatively small number of
 particles are excited above the Fermi surface leaving an
 unequal amount of holes below the Fermi surface due to the collective
 effect. At high temperatures, there are a large number of holes which
 drive the system towards a Maxwell-Boltzman distribution of particles.

\section{Distribution profiles}
\label{distribution}
 
A quantitative explanation of the GES induced by the dynamical interaction $\gamma$ and the anyonic
statistical interaction $\kappa$ follows from the calculation of the
statistics and the thermodynamics via either the TBA equation  (\ref{TBA}) or
the GES results (\ref{FS}) and  (\ref{FS-2}).  
To this end, we express the particle number and total energy as
\begin{eqnarray}
N=\int_0^{\infty}d\epsilon\, G(\epsilon) n(\epsilon),\quad 
E=\int_0^{\infty}d\epsilon \,G(\epsilon) n(\epsilon)\epsilon 
\end{eqnarray}
where $n(\epsilon)$ is determined from (\ref{FS}). 
The pressure is given by the relation $PL=2E$ and the 
density of states by 
\begin{equation}
 G(\epsilon)=L/\left(2\pi
\sqrt{\frac{\hbar^2}{2m}\epsilon }\right).
\end{equation}  
In the thermodynamic limit we may ignore the factor $\nu/L$, as it is 
much smaller than the cut-off quasimomentum. 

In order to calculate the thermodynamic properties at low
temperatures, we bring them into line with the Sommerfeld expansion. 
For convenience, we make the change $w(\epsilon)\to
1/w(\epsilon)-1$ in equations (\ref{FS}) and (\ref{FS-2}), giving 
\begin{eqnarray}
& &n(\epsilon)=\frac{w(\epsilon)}{1+(\alpha-1) w(\epsilon)},
\label{FS-new}\\
& &\alpha\ln \left(1-w(\epsilon)\right)-\ln
w(\epsilon)=\frac{\epsilon-\mu}{K_BT}. \label{FS-2-new}
\end{eqnarray}
We find that $w(\infty) \approx 0$ and 
\begin{equation}
w(0)\approx 1-{\mathrm
e}^{-\frac{\mu}{\alpha K_BT}}-\frac{1}{\alpha}{\mathrm
e}^{-\frac{2\mu}{\alpha K_BT}}+O({\mathrm e}^{-\frac{3\mu}{\alpha
K_BT}}).
\end{equation}
Following Isakov {\em et al.} \cite{Isakov2}, at low
temperatures, i.e., for $T < T_d$, where $T_d=
\frac{\hbar^2}{2m}n^2$ is the quantum degeneracy temperature, the
thermodynamics of ideal particles obeying GES can be derived from the
fractional statistics $n(\epsilon)$  (\ref{FS-new}) via Sommerfeld
expansion. Let us define $I[f] = \int_{0}^{\infty}d\epsilon
f(\epsilon)n(\epsilon)$.
This integral can be calculated explicitly with the help of
(\ref{FS-new}) and (\ref{FS-2-new}). It follows that
\begin{equation}
I[f]=\frac{1}{\alpha}\int_0^{\mu}d\epsilon f(u)+\sum_{l=1}^{\infty}\frac{(K_BT)^{l+1}}{l!}C_l(\alpha)f^{(l)}(u),
\label{GES-I}
\end{equation}
where the coefficents are given by
\begin{equation}
C_l(\alpha)=\sum_{k=0}^{l-1}\alpha ^k(-1)^{l-k}\left(\begin{array}{l}
  l\\k\end{array}\right)\int_0^1\frac{d w}{w}\ln ^k w \ln^{l-k} (1-w).
\end{equation}
Here $f^{(l)}(u)$ indicates the $l$th
order derivative of the function $f(u)$ with respect to $u$ in the usual way.

{}From the relation
\begin{equation}
\frac{N}{L}=\frac{1}{2\pi
  \sqrt{\frac{\hbar^2}{2m}}}\int_0^{\infty}d\epsilon \,
  n(\epsilon)\epsilon^{-\frac{1}{2}}
\end{equation}
we find the chemical potential
\begin{equation}
\mu
=\mu_0\left[1+c_2t^2+c_3t^3+c_4t^4+O(t^5)\right] \label{GES-mu}
\end{equation}
where the coefficients are given by
\begin{eqnarray}
c_2 &=& \frac{\pi^2\alpha}{12}, \qquad c_3 =
-\frac{3}{4}\zeta(3)\alpha(1-\alpha),\nonumber\\
c_4 &=& \frac{\pi^4}{144}\alpha\left(3-2\alpha+3\alpha^2\right), 
\end{eqnarray}
with $\zeta(3)=\sum_{n=1}^{\infty}1/n^3$. 
In the above we introduced an effective temperature $t={K_BT}/{\mu_0}$,
where $\mu_0=\frac{\hbar^2}{2m}k_F^2\alpha ^2$ is the anyon cut-off energy
at $T=0$ and $k_F =n\pi$ is the Fermi momentum. 
The result (\ref{GES-mu}) reduces to the chemical potential for the 1D free Fermi gas
in the limit $\gamma \to \infty$.

\begin{center}
\begin{figure}
\includegraphics[width=0.98\linewidth]{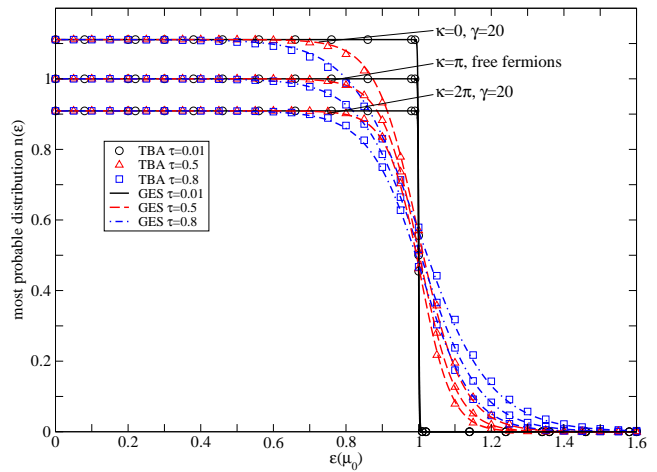}
\caption{Comparison between the most probable distribution profiles $n(\epsilon)$ for the values 
  $\gamma =20$ and $\kappa=0,\, \pi,\, 2\pi$ at different values of the degeneracy temperature
  $\tau=K_BT/T_d$.  At
  zero temperature $n(\epsilon)={1}/{\alpha}$ leads to an anyon
  surface at $\epsilon =\mu_0$.  At finite temperatures these surfaces
  are gradually diminished by a large number of holes below the
  surface. Pure Fermi-Dirac statistics appear for $\gamma \to
  \infty$ or $\kappa=\pi,3\pi$.  For temperature $\tau<0.1$,
  thermal fluctuations are suppressed by the dynamical interaction. The
   lines show the most probable GES 
  distribution derived from (\ref{FS-new}) with (\ref{GES-mu}). The
  symbols show the corresponding
  distributions evaluated from the TBA result (\ref{TBA-n}) for
  interacting anyons.  
}
\label{fig:n}
\end{figure}
\end{center}

At finite temperatures, the deviations of the distributions for strongly interacting anyons from Fermi-Dirac statistics 
are induced by particle excitations which leave holes below the Fermi surface.
It is clearly seen that the dynamical interaction $\gamma$ and the
anyonic statistical interaction $\kappa$ continuously vary the GES,
with the most probable distribution of anyons with nonmutual GES
approaching free fermion statistics as the interaction increases.
Distribution profiles obtained from the GES results (\ref{FS}) and (\ref{GES-mu}) 
are shown in Figure \ref{fig:n}.

Some analytic expressions  for the  distribution function may be obtained in terms of the variable 
$x = {\mathrm e}^{\frac{\epsilon-\mu}{K_BT}}$, 
namely 
\begin{equation}
n(\epsilon)=\frac{1}{\alpha}\sum_{m=0}^{\infty}d_mx^{\frac{m}{\alpha}},
\end{equation} 
which is seen to hold for $\epsilon < 0.9 \mu_0$. 
The coefficients $d_m$ are $d_0=1$ and 
\begin{equation}
d_m=\frac{(-1)^m}{m!\alpha^m}\prod_{k=0}^{m-1}(m-\alpha k)
\end{equation} 
for $m=1,\ldots, 4$. 
We expect that this expression may be exact for all $m$.
On the other hand, for $\epsilon> 1.1 \mu_0$, we have a universal
relation $n(\epsilon)=\sum_{m=1}^{\infty}Q_mx^{-m}$ \cite{Isakov2}, where
\begin{equation}
Q_m=\prod^{m-1}_{k=1}\left(1-\alpha\frac{m}{k}\right).  
\end{equation}

All distribution curves pass through the inflection point $1/(2\alpha)$. 
The interactions $\gamma$ and $\kappa$ are seen to vary the height of the plateaux. 
This change of height with respect to $\gamma$ has been observed in the experimental momentum 
distribution profiles for the lattice TG gas \cite{T-G1}. 
For $\kappa=\pi -\theta $ and $\kappa=\pi+\theta $, the anyonic TG gas and the super anyonic TG gas form a 
GES duality between $\alpha$ and $1/\alpha$, for example, for $\kappa=0$ and $\kappa =2\pi$.

On the other hand, for the strong coupling limit, the distribution
function obtained from the TBA (\ref{TBA}) and (\ref{d-BA}) is 
\begin{equation}
n(\epsilon)=\frac{1}{\alpha(1+{\mathrm e}^{\frac{(\epsilon-\mu)}{K_BT}})} \label{TBA-n}
\end{equation}
with $\epsilon=\frac{\hbar^2}{2m}k^2$. 
After some algebra, we find that
\begin{eqnarray}
I[f]&=&\int_0^{\infty}d\epsilon f(\epsilon)n(\epsilon)\nonumber\\
    &=& \frac{1}{\alpha}\left(\int_0^{\mu} f(\epsilon)d\epsilon
    +\frac{\pi^2}{6}(K_BT)^2f^{(1)}(u) \right.\nonumber \\
& &\left.+\frac{7\pi^4}{
    360}(K_BT)^4f^{(3)}(u)+\frac{31\pi^6}{126 \times 5!
    }(K_BT)^6f^{(5)}(u)\right.\nonumber \\
& &\left. +\frac{127\pi^8}{120 \times 7!
    }(K_BT)^8f^{(7)}(u) +O((K_BT)^{10}) \right). \label{TBA-Ite}
\end{eqnarray}
It follows that  
\begin{equation}
\mu =\mu_0\left[1+\frac{\pi^2}{12}t^2+\frac{\pi^4}{36}t^4 +O(t^6)\right]
\end{equation}
with  $\mu_0=\frac{\hbar^2}{2m}k_F^2\alpha ^2$.
Figure \ref{fig:n} also shows the close agreement between the profile (\ref{TBA-n}) of strongly
interacting anyons derived from the TBA and the most probable distribution (\ref{FS}) 
for ideal particles obeying nonmutual GES.
In general both results are seen to coincide well at low temperatures in the strongly
interacting regime. 

\section{Thermodynamics and local correlations}
\label{TD}

Thermodynamic properties, such as the pressure $P$ and the total energy
$E$, may be calculated directly from the integral $I[f]$ given in Ref.
\cite{Isakov2} for ideal particles obeying GES on the one hand 
and via the TBA result (\ref{TBA-Ite}) on the other.  
The pressure and total energy following from the GES result (\ref{FS}) and (\ref{GES-I}) are
\begin{eqnarray}
P&=&\frac{2}{3}n\mu_0\left[1+3c_2t^2+2c_3t^3+\frac{9}{5}c_4t^4+O(t^5)\right],~~\label{GES-P}\\
E&=& E_0\left[1+3c_2t^2+2c_3t^3+\frac{9}{5}c_4t^4+O(t^5)\right]. \label{GES-E}
\end{eqnarray}
Here $E_0=\frac{1}{3}N\mu_0$ is the ground state energy at zero
temperature \cite{BGO}.
On the other hand, the pressure and total energy obtained from the TBA result (\ref{TBA-Ite}) are
\begin{eqnarray}
P&=&\frac{2}{3}n\mu_0\left[1+\frac{\pi^2}{4}t^2+\frac{\pi^4}{20}t^4+O(t^6) \right],\label{TBA-P}\\
E&=&E_0\left[1+\frac{\pi^2}{4}t^2+\frac{\pi^4}{20}t^4+O(t^6) \right].\label{TBA-E}
\end{eqnarray}
The TBA  results coincide with the corresponding GES results (\ref{GES-P}) and (\ref{GES-E}) in 
the TG regime at low temperatures, i.e., for large $\gamma$ and $T\ll T_d$.

\begin{center}
\begin{figure}
\includegraphics[width=0.85\linewidth]{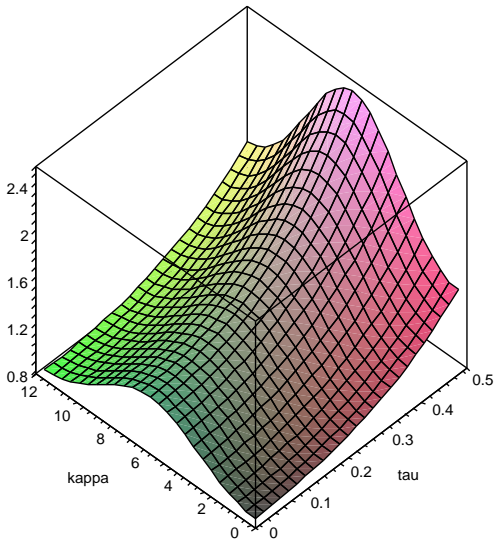} 
\caption{The energy per particle in units of the free Fermi energy $E_{\rm F}$ vs the
  effective temperature $\tau=K_BT/T_d$ and the anyonic parameter $\kappa$. 
  Here the interaction strength  is fixed to be $\gamma =20$ in
  (\ref{GES-E}) in the strong coupling regime.  For $0 < \kappa < \pi$ the energy curves interpolate 
  between strongly interacting bosons at $\kappa = 0$ and free fermions at $\kappa=\pi$.
For the super TG anyons where  $\pi < \kappa <  3\pi$,  the energy is greater than the energy for free fermions.
For  $3\pi < \kappa <  4\pi$ the interpolation is from free fermions
to interacting bosons.}
\label{fig:E}
\end{figure}
\end{center}

At low temperature $T\ll T_d$, the leading terms in the thermodynamic
properties reveal the signature of the degenerate anyon gas.  
For the super anyonic TG gas with $\pi< \kappa < 3\pi $ the total energy is larger than the free Fermi energy
$E_{\rm F}=\frac{1}{3}N\left(\frac{\hbar^2}{2m}k_F^2\right)$.  
In particular, the anyonic parameter $\kappa$ varies the GES parameter  from
$\alpha <1$ to $\alpha >1$, i.e., it may trigger a phase beyond the free Fermi gas.  
Fig.~\ref{fig:E} shows that the energy (\ref{GES-E})
increases as the effective temperature increases. 
We see clearly from Fig.~\ref{fig:E} that the anyonic parameter triggers 
a phase beyond the free Fermi gas for $\pi < \kappa< 3\pi$.

In the grand canonical description, the free energy per particle $f$ follows from 
$f=F/N=\mu -PL/N$.
Thus the  specific heat is given by
\begin{equation}
c_v=-T\left(\frac{\partial^2 F}{\partial T^2}
\right)_{N,L}=\left(\frac{\partial E}{\partial T}\right)_{N,L}
\end{equation}
with result
\begin{eqnarray}
c_v=\frac{NK_B\tau }{6\alpha }\left[1-\frac{9(1-\alpha
  )\zeta(3)}{\pi^4\alpha^2}\tau +\frac{(3-2\alpha+3\alpha^2)}{10\pi^2\alpha^4}\tau^2\right].\nonumber\\
  \label{Cv}
\end{eqnarray}

\begin{center}
\begin{figure}
\vskip -7mm
\includegraphics[width=0.90\linewidth]{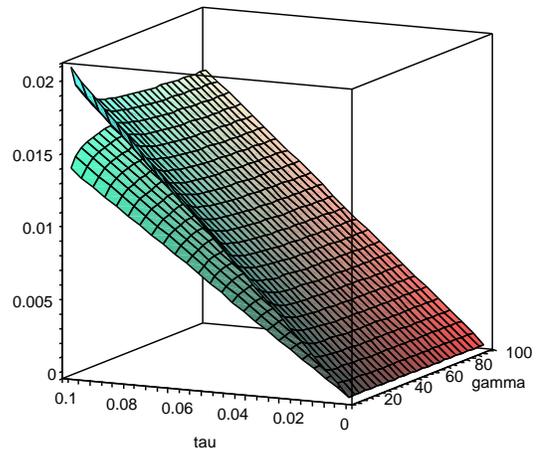}
\vskip -5mm
\caption{Specific heat (\ref{Cv}) in units of $NK_B$ vs the degeneracy temperature
  $\tau=K_BT/T_d$ and the interaction parameter $\gamma$. The upper layer is the 
  TG Bose gas with $\kappa=0$, while the lower layer is the super anyonic TG
  gas with $\kappa=2\pi$. The specific heat decreases with increasing
  interaction for the TG Bose gas while it increases with
  increasing interaction for the super anyonic TG gas. They form a GES
  duality between $\alpha$ and $1/\alpha$ in the strong coupling limit. For both cases
  the specific heat is almost linearly increasing with the temperature. }
\label{fig:Cv-L}
\end{figure}
\end{center}

For $\gamma \to \infty$, the second term in $c_v$ vanishes, recovering the result for the free Fermi gas, 
up to irrelevant corrections at low temperatures. 
Here the TBA gives the result 
\begin{equation}
c_v=\frac{NK_B\tau }{6\alpha
  ^2}\left(1-\frac{4}{10\pi^2\alpha^4}\tau^2\right).
\end{equation}
We plot the specific heat in units of $NK_B$ against the degeneracy
temperature $\tau=K_BT/T_d$ and interaction strength $\gamma$ in
Fig. \ref{fig:Cv-L}.
The specific heat almost linearly increases as the temperature
increases. However, for strongly coupled anyons $c_v$ deviates upwards
from the free Fermi curve as $\gamma$ becomes weaker.  This is mainly
because the dynamical interaction $\gamma$ lowers the entropy in
fermionization. While for the super anyonic TG gas, $c_v$ deviates
downwards from the free Fermi curve as $\gamma$ decreases.  This kind
of GES-dependent specific heat is also seen for the three-dimensional ideal
anyon gas \cite{Isakov,Aoyama}. In general the specific heat reveals an important
signature of quantum statistics for interacting many-body systems.

The local two-particle correlations can be used to
classify various finite temperature regimes and study phase coherence
behaviour in 1D interacting quantum  gases \cite{Shlyapnikov}. 
In a significant advance, the local pair correlations for the 1D Bose gas have been observed 
experimentally by measuring photoassociation rates \cite{Weiss}.  
We now consider the local pair correlations for the 1D anyon gas and their
role in revealing GES. 
In the grand canonical description, the two-particle local
correlations are given by \cite{Shlyapnikov}
\begin{equation}
g^{(2)}(0)=\frac{2m}{\hbar^2n^2}\left(\frac{\partial
  f(\gamma,t)}{\partial \gamma }\right)_{n,t}.
\end{equation}  
Using the GES result (\ref{FS}) we obtain the two-particle local correlations at low
temperatures
\begin{equation}
g^{(2)}(0)\approx \frac{4\pi^2\cos(\frac{\kappa}{2})}{3\gamma^2}\left(1+\frac{\tau^2}{8\pi^2}
-\frac{3\zeta(3)\tau^3}{8\pi^6} +\frac{\tau^4}{480\pi^4}\right)
\end{equation}
which is to be compared with the TBA result
\begin{equation}
g^{(2)}(0)\approx
\frac{4\pi^2\cos(\frac{\kappa}{2})}{3\gamma^2}\left(1+\frac{\tau^2}{4\pi^2}
+\frac{3\tau^4}{80\pi^4}\right).
\end{equation}
For $\kappa=0$ the latter coincides with the result for the 1D interacting Bose gas \cite{Shlyapnikov}.
We see that the dynamical interaction dramatically reduces the pair correlations 
due to decoherence between individual wave functions. 
The local pair correlations increase with increasing temperature.  
At temperatures $T \ll T_d$, the local pair correlations approach free Fermi behaviour 
as $\gamma \to \infty$ or $\kappa\to \pi$.  
This quantitatively agrees with the experimental observation for 1D interacting bosons \cite{Weiss}.

\section{Conclusion}
\label{conclusion}

To conclude we have demonstrated that the statistical signature of the strongly interacting anyon gas 
is equivalent  at low temperatures to a gas of ideal particles obeying nonmutual GES. 
Through GES and the TBA, we have independently derived and compared analytic results for the  
thermodynamics and local pair correlations.  
Both the dynamical interaction and the anyonic statistical interaction 
implement a continuous range of GES.
They implement a duality between the GES parameter $\alpha$ and $1/\alpha$ and trigger 
interesting degenerate behaviour at low temperatures. 
It is shown that the statistical profiles of quasiparticles for the 
$1$D interacting model of anyons are influenced by both the dynamical
interaction and anyonic statistical interaction. 
More generally it will be interesting to see how the statistical properties of 
degenerate systems are affected by dimensionality, dynamical
interaction, topological effects and internal degrees of freedom.
In future we hope to address how internal spin degrees affect
the statistical profiles of interacting particles \cite{Fukui,Iguchi2}.  
The statistical profiles reveal  an important collective signature of
interacting quantum degenerate gases.
This suggests the possibility of observing the quantum degenerate
behaviour of ideal particles obeying GES through experiments on 1D
interacting systems.


\noindent
{\em Acknowledgements.}  The authors acknowledge the Australian Research Council for support. 
They also thank M. Bortz for discussions on the TBA results, N. Oelkers and C.H. Lee for assistance 
with the figures and Z. Tsuboi for pointing out Refs.~\cite{Iguchi,Iguchi2}.


\end{document}